\documentclass[12pt,a4paper]{article}
\usepackage[T2A]{fontenc}
\usepackage[cp1251]{inputenc}
\usepackage[english]{babel}
\usepackage{amssymb}
\usepackage{amsmath}
\usepackage{graphicx}
\usepackage[small,sc,center]{titlesec}
\usepackage{indentfirst}
\usepackage[labelsep=period]{caption}
\usepackage{caption}
\newcommand{\msun}{\,$M_{\odot}$}
\newcommand{\msyr}{\,$M_{\odot}$\,yr$^{-1}$}

\newcommand{\ergs}{\,erg\,s$^{-1}$}

\newcommand{\gcm}{\,g\,\,cm$^{-1}$}
\newcommand{\gcmsq}{\,g\,cm$^{-2}$}
\newcommand{\kms}{\,km\,s$^{-1}$}

\newcommand{\cmq}{\,cm$^{-3}$}
\newcommand{\cmsqg}{\,cm$^2$\,g$^{-1}$}
\newcommand{\ha}{H$\alpha$}
\newcommand{\heii}{He\,II}
\newcommand{\ovi}{O\,VI}
\begin{document}
	
\begin{center}
\textbf{\large Origin of broad \heii\,4686\,\AA\ emission in early spectra of type IIP supernovae}
\vskip 5mm
\copyright\quad
2024 \quad N. N. Chugai\footnote{email: nchugai@inasan.ru} and V. P. Utrobin$^{2,1}$\\
\textit{$^1$ Institute of astronomy, Russian Academy of Sciences, Moscow} \\
\textit{$^2$SRC ``Kurchatov institute'', Moscow} \\
Submitted  05.10.2023 
\end{center}

{\em Keywords:\/} stars -- supernovae; stars -- stellar wind

\noindent
{\em PACS codes:\/} 

\clearpage
 
 \begin{abstract} 
 	
 We propose a model for the origin of the broad \heii\,4686\,\AA\ emission in the early spectrum
 of type II SN~2020jfo.
The 4686\AA\ line is emitted presumably by dense fragments embedded into a hot gas of the    
  forward shock wave. 	
The fragments are produced as a result of a heavy braking of the dense low-mass 
  shell at the ejecta boundary and a simultaneous Rayleigh-Taylor instability.
The temperature of line-emitting fragments is  $\approx$5$\times10^4$\,K.
Calculations of ionization and excitation of helium and hydrogen account for the 
 \heii\,4686\,\AA\ luminosity, the large flux ratio of  \heii\,4686\,\AA/\ha, and a significant   
 optical depth of the 4686\,\AA\ line.
We demonstrate that fragments heating by hot electrons behind the forward shock compensates 
  cooling via the \heii\,304\,\AA\ emission.

\end{abstract}

\section{Introduction}

Supernovae (SNe) of types IIP and IIL are the result of a core collapse of a massive 
 red supergiant (RSG). 
In some cases early spectra show signature of more vigorous mass loss by a presupernova 
 (pre-SN) compared to ordinary RSG (Chugai 2001; Groh 2014; Yaron et al. 2017).
The density of the circumstellar (CS) matter in a close vicinity is a crucial factor that 
 can affect the SN bolometric luminosity and this effect should be taken into account 
  in the hydrodynamic modelling (Blinnikov \& Bartunov 1993; Chugai 2001; Morozova et al. 2017).

Despite a high degree of development and verification of applied methods of the radiation  
  hydrodynamics, the SN parameters (explosion energy, ejecta mass and pre-SN radius) 
  recovered by different authors can differ by a factor of $1.5-2$ due to a different choice of the CS gas density.
This situation is illustrated by type IIP SN~2020jfo: the CS shell mass within the radius   
 of $10^{15}$\,cm is 0.2\msun\ in the model by Teja et al. (2022) compared to $\sim$10$^{-3}$\msun\ 
 in the alternative model (Utrobin \& Chugai 2024).
In the first case the large CS mass is obtained based on the description of the early stage 
of the light curve, whereas in the second case the mass of the CS shell is the result 
of both the hydrodynamic modelling and making use of the spectral information, particularly, on the \heii\ 4686\,\AA\ emission.
The latter indicates the high expansion velocity, i.e., the negligible deceleration and   therefore, the rarefied  CS medium.

The \heii\ 4686\,\AA\ emission is an essentially unique noticeable broad line in early spectrum, which
 permits us to directly estimate the expansion velocity of the outermost ejecta of SN~IIP/L at the very early stage.
Realizing a diagnostic value of this line, it is appropriate to pose a question, whether we 
  understand the physical conditions in the line-emitting region adequately enough to rule out doubts on the boundary velocity recovered from this line.
 
The broad \heii\,4686\AA\ emission was observed during the initial several days after the explosion 
  in other SNe~IIP as well, including SN~2006bp (Quimby et al. 2007), SN~2013fs (Bullivant et al. 2018), SN~2017gmr (Andrews et al. 2019), and SN~2023ixf (Jacobson-Gal\'{a}n et al. 2023).
In each case the \heii\ 4686\AA\ indicates large expansion velocity of the line-emitting gas 
 and significant blueshift.
Another important feature is a high flux ratio $f(4686)/f(\mbox{H}\alpha)$.
For SN~2020jfo we estimate 
 this ratio in the range of $4-5$  based on the spectrum of Teja et al. (2022).
This can be compared to the maximum ratio among planetary nebulae $f(4686)/f(\mbox{H}\alpha) \sim 0.2$  (Bohigas 2022).

There is no common opinion on the origin of the broad 4686\,\AA\ emission. 
A model proposed earlier for the 4686\,\AA\ line in SN~2013fs suggests 
 that this line is emitted by dense fragments produced as a result of the deceleration of 
  outer SN layers and a concomitant Rayleigh-Taylor (RT) instability (Chugai 2020).
However, the issue of physical conditions in the line-emitting fragments was not discussed, 
 except for the conclusion that the layer of fragments must be thin and adjoin the photosphere  with sharp boundary.
Besides, the dome-like profile shape suggests that fragments should be optically thick in 
 the 4686\,\AA\ line frequencies. 

Recently Shrestha et al. (2023) have compared the \heii\,4686\,\AA\ in the spectrum of SN 2023axu with 
 the model r1w1 of early spectrum of SN~IIP with a CS wind of $\dot{M} = 10^{-6}(u/10\mbox{km/s})$\msyr\ (Dessart et al. 2017).
The model satisfactorily describes the 4686\,\AA\ profile at +1.1\,d, but at later moment 
 (+1.5\,d) the agreement is broken: the model profile becomes symmetric in contrast  
   to the blueshift and asymmetry of the observed profile. 
It should be emphasized that in the model r1w1 the adiabatic forward shock is absent despite 
 the low wind density.   
The issue of the adequate model for the broad \heii\,4686\,\AA\ emission in early spectra of SNe~IIP thus remains open.

Below we study conditions that account for the major properties of the \\
broad \heii\,4686\,\AA\ emission in the spectrum of SN~2020jfo.
We start with a general picture, then calculate the luminosities of the \ha\ and 4686\,\AA\ lines,
 and recover conditions (mass of emitting fragments and kinetic temperature), which 
   reproduce the major properties of the 4686\,\AA\ emission.
Finally, we consider the issue of fragments heating by hot electrons of the forward shock, which 
 should compensate the cooling via the \heii\,304\,\AA\ line.

%================================================================
\begin{figure}
	\centering
	\includegraphics[width=0.5\columnwidth]{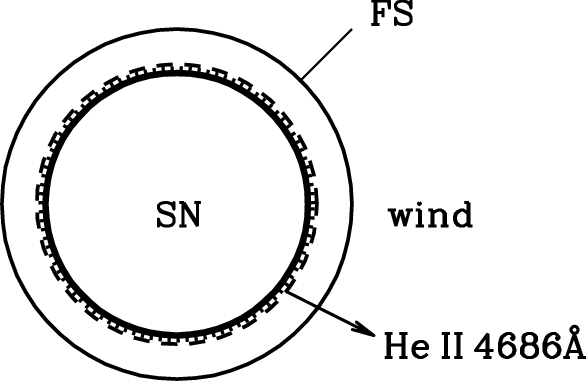}
	\caption{
		Schematic picture for the origin of the \heii\,4686\,\AA\ emission.
		Unperturbed SN envelope is bounded by the thin dense shell that serves as 
		the photosphere (thick line circle).
		Right outside the photosphere lies the layer of dense fragments produced due to the RT instability 
		brought about by the deceleration of outermost ejecta in the wind. Hot gas of the adiabatic forward shock
		maintains the high density and heating of fragments responsible for the \heii\,4686\,\AA\ emission.
	}
\label{fig:cart}
\end{figure}
%==================================================================

\section{Conditions in 4686\AA\ line-emitting zone}

\subsection{General picture}

According to the hydrodynamic model of SN~2020jfo, the shock breakout (SBO) occurs 
 at 0.5\,d after the explosion (Utrobin \& Chugai 2024).
Following the SBO, the SN ejecta generates a forward shock wave in the pre-SN wind that, for the moderate wind density,  turns out adiabatic in contrast to the radiative reverse shock.
However, at the very early stage the viscous velocity jump at the forward shock can be small compared to 
  the boundary SN velocity because of the wind acceleration by the SN radiation. 
 Let us estimate this effect. 

In SN~2020jfo at $t = 2$\,d the photosphere with the luminosity of
 $L \approx 3\times10^{42}$\ergs\ 
coincides with the thin shell at the SN boundary 
that expands with the speed of 16500\kms.
The SN radiation accelerates the preshock wind up to a velocity 
\begin{equation}
v_{acc} \approx \frac{\kappa L}{4\pi v^2ct}	\approx  10^3L_{43}/v_4^2t_d \quad \mbox{km\,s}^{-1},
\end{equation}	
  where $\kappa = 0.34$\cmsqg\ is the Thomson opacity, $v$ is the SN boundary velocity, 
   $c$ is speed of light, 
  $L_{43}$ is the SN luminosity in $10^{43}$\ergs, $v_4$ is the SN boundary velocity in $10^4$\kms,
  $t_d$ is time in days. 
 For $v_4 =1.65$ and $t_d =2$ the preshock wind velocity is about 200\kms, almost two order lower 
  the boundary SN speed.
The contribution of a line opacity could increase this estimate by a factor less than two 
 (Chugai et al. 2002).

We therefore conclude that the radiative acceleration of the preshock wind does not 
 affect the viscous jump and the temperature at the forward shock.
This conclusion is not relevant in the case of a massive optically thick CS shell around some 
 SNe, when the forward shock due to a strong preshock heating by the radiation presursor 
 (aka Marshak wave) propagates in the regime of the isothermal jump (Blinnikov 2008).

Noteworthy, after the SBO the formation of the adiabatic forward shock in the rarefied wind,
  $w =  \dot{M}/u \lesssim 10^{15}$\gcm, poses problems for the hydrodynamic modelling.
This is related to the rapid transition from the optically thick to the optically thin regime 
 of the shock wave propagation. 
Probably for this reason the SN~IIP explosion and a subsequent formation of the adiabatic forward  shock with the temperature of $\sim$100\,keV has not yet been implemented.
The early formation of the adiabatic forward shock in SN~IIP is evidenced by the 
  hard X-rays ($T_x > 60$\,keV) from SN~2023ixf on day 4 after the explosion 
   (Grefenstette et al. 2023), despite 
 the CS wind in this case is relatively dense,  $w \sim 3\times10^{15}$\gcm.
  
The broad  \heii\,4686\,A\ emission is present in the spectrum of SN~2020jfo (Teja et al. 2022)
  on day 2.1 after the explosion (the adopted explosion date is MJD 58973.83) 
   with the luminosity of $\approx$4$\times10^{39}$\ergs.
According to the SN hydrodynamic model  and the CS interaction model (Utrobin \& Chugai 2024), at this moment the SN boundary velocity is 16500\kms and the radius 
 is  $3.2\times10^{14}$\,cm.
The inferred wind parameter $w = \dot{M}/u = 2.2\times10^{15}$\gcm\ ($u$ is the wind speed)
 suggests the baryon density $n_w = 10^9$\cmq at this radius.

The outer ejecta boundary serves as a spherical piston  that drives the forward shock with the 
 postshock temperature $T_s = (3/16)\mu m_p v_s^2/k = \\ 
 4\times10^9(v_s/17000\,\mbox{km/s})^2$\,K   and density $n_s = 4n_w$. 
The forward shock at the given moment is essentially adiabatic.
Interestingly, the Compton cooling of hot electrons dominates over the bremsstrahlung radiation but this does not change conclusion on the adiabatic regime of the forward shock.     

In the proposed picture the \heii\ 4686\AA\ emission forms in the narrow layer right next to 
 the photosphere that in turn coincides with the boundary ejecta dense shell (Fig.~1).
The line-emitting layer is a two-phase medium in which the relatively cold dense fragments with 
 temperature  $T \lesssim 10^5$\,K are embedded in the hot gas with temperature 
  $T_s \approx 4\times10^9$\,K.
The fragments are presumably produced as a result of the RT instability 
  developed due to the deceleration of the outermost ejecta in the CS gas.

In the development of the RT instability of a decelerated dense shell one can distinguish three stages (cf. Blondin \& Ellison 2001). 
At the first stage spikes of the cold dense shell penetrate the  rarefied hot gas of the forward shock.
At the second stage spikes are  subject to the Kelvin-Helmholtz (KH) instability that generates a mushroom structure of spikes.
This brings about the formation of thin layers of the cold dense gas with a large surface area. 
These dense layers are presumably responsible for the emergence of the \heii\ 4686\AA\ emission.
The third stage corresponds to a final fragmentation and a complete mixing of the dense cold gas with the  hot gas of the forward shock.
The fragments mixing is signaled by the disappearance of the broad  \heii\ 4686\AA\ emission.

The scenario based on the RT instability of the boundary dense shell has an important advantage compared to the conceivable origin of the \heii\ 4686\AA\ emission 
  from a stable thin dense shell, because
 it connects the duration of the \heii\ 4686\AA\ emission ($2-3$\,d) with a total time of the development 
 of the RT instability until the complete mixing of cold fragments with the hot gas.
This chain of events is inconsistent with the long-duration emission from the stable thin dense  shell heated by the hard radiation or hot electrons of the forward shock.
%In turn, the long-duration emission from the stable thin dense shell heated by
%   the hard radiation or hot electrons of the forward shock is inconsistent with
%   the characteristic duration of the \heii\ 4686\AA\ emission.

The noted unusually high flux ratio \mbox{$f(4686)/f(\mbox{H}\alpha)\sim 4-5$} in the \\
  SN~2020jfo spectrum indicates a dominant role of the collisional excitation of 
  \heii. 
This, in turn, suggests the kinetic temperature close to the temperature of the collisional ionization of \heii, i.e. $5\times10^4$\,K.  
Hydrogen in this case is strongly ionized and its emission is predominantly due to  the recombination with a 
  minor contribution of collisional excitation because of low fraction of neutrals.
That presumably is the physics behind  the large flux ratio  \heii\,4686\AA/\ha.

%========================================================
\begin{table}[t]
	\vspace{6mm}
	\centering
	{{\bf Table.  } Model parameters. }\\
	
	\bigskip	 
	
	\begin{tabular}{l|c} 
		\hline
\noalign{\smallskip}       
		Parameter     & Value  \\
\noalign{\smallskip}       
		\hline	
\noalign{\smallskip}       
%		&       \\	
		Radius $r$ [$10^{14}$\,cm]	   &    3.2 \\
		Velocity $v$ [$10^9$\,cm/s]      &  1.65 \\
		Wind parameter $w$ [$10^{15}$\,g\,cm$^{-1}$] & 2.2\\
\noalign{\smallskip}       
		\hline
\noalign{\smallskip}       
%		&       \\
		Mass of cold component  $M$ [\msun] & $10^{-8}$ \\
		Parameter of surface area $\zeta$            &  2.5 \\
	   Fragments surface density $N_b$ [cm$^{-2}$]   &  $3.7\times10^{18}$\\
\noalign{\smallskip}       
		\hline
	\end{tabular}
\end{table}
%=======================================================

\subsection{Ionization, excitation and fragments emission} 

To confirm the outlined picture, we calculate ionization and excitation of  
  \heii\ and hydrogen 
 for certain mass of fragments and kinetic temperature in the range of $10^4 - 10^6$\,K.
In the isobaric approximation the baryon number density in a fragment for a given   
  temperature of cold gas $T$ is  equal $n = n_sT_s/T$. 
The emitting gas mass and density imply an emitting volume and a surface baryon density of fragments 
  $N_b = M/4\pi r^2\zeta m_p$, where $r$ is the radius of the boundary dense shell, and $\zeta$ 
 is the parameter of the total surface area of fragments --- one sided surface of flat fragments.
The latter parameter, generally, might be found from 3D hydrodynamic modelling of the SN/CSM interaction and the concomitant RT instability.
Yet the complexity of this approach is obvious even in the  adiabatic case (Blondin \& Ellison 2001).

The ionization fractions of hydrogen and helium for the normal abundances are  
  calculated with the collisional ionization from two low levels and the radiative 
  Case B recombination.
A photoionization by the SN radiation is neglected.
A population of the second level of hydrogen and \heii\ is determined taking into 
  account  a multiple scattering of the resonant radiation and a local escape 
   with the probability  
$\beta_{ik} = [1 - \exp(-\tau_{ik})]/\tau_{ik}$, where  
 $\tau_{ik} = \sigma_0f_{ik}\lambda_{ik}N_i/u_{t}$.
 Here $\sigma_0$ is the line integral cross section, $f_{ik}$ is the oscillator strength, $\lambda_{ik}$ is the wavelength, $N_i$ is the column density of ion/atom on the lower 
 level of a transition, $u_t$ is the local velocity dispersion adopted to be equal the isothermal sound  velocity.

The collisional ionization of hydrogen and helium is calculated using the classical ionization cross section averaged over the Maxwell distribution.
The collisional excitation rates for \heii\ are determined in the van Regemorter (1962) 
approximation, whereas for hydrogen we use the collisional rates from Vernazza et al. (1981). 
For hydrogen we take into account three lower levels plus continuum, whereas for \heii\ four lower levels plus continuum  are taken into account.

In the local escape approximation the line luminosity is determined by the total rate of the 
  collisional excitation of the upper level of a certain transition $C_k$ [cm$^{-3}$\,s$^{-1}$], by the rate 
  of the local escape $A_{ki}\beta_{ik}$, and by the rate of the collisional transition to all  lower levels $D_k$. The line luminosity is then 
$L_{ik} = C_kh\nu_{ik} VA_{ki}\beta_{ik}/(A_{ki}\beta_{ik} + D_k)$, where $V$ is the total volume 
of the line-emitting fragments.
Recombination emission rates for the \ha\ and \heii\ lines correspond to the 
Case B (Osterbrock \& Ferland 2006).

%================================================================
\begin{figure}
	\centering
	\includegraphics[width=0.9\columnwidth]{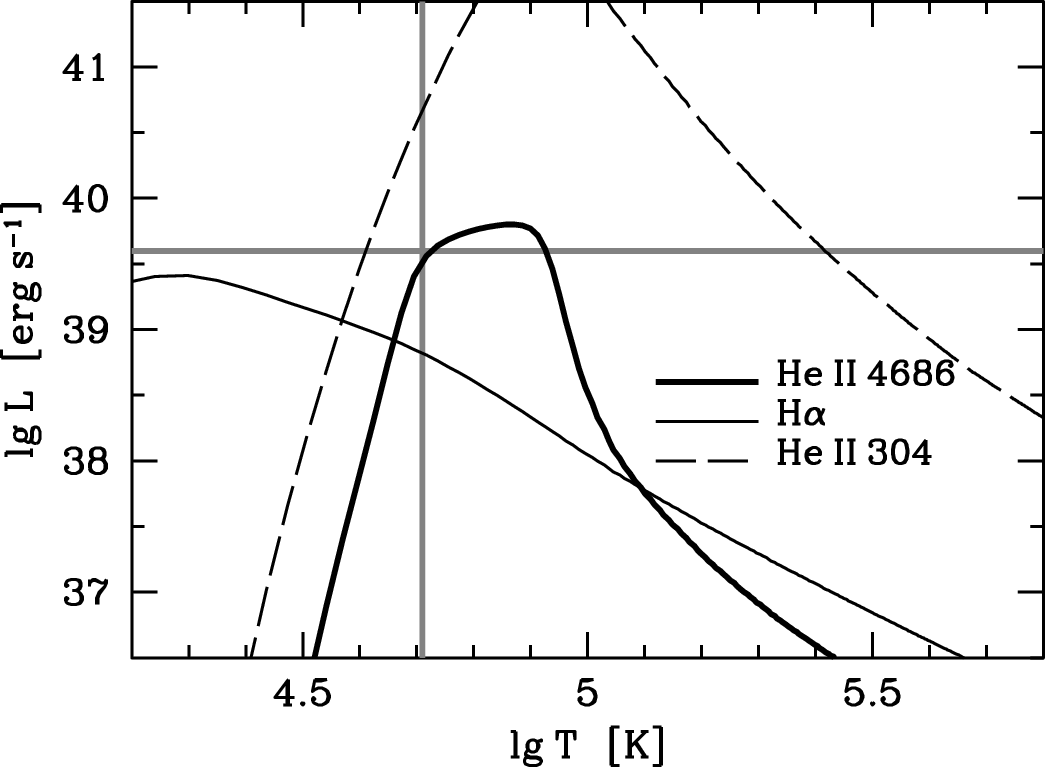}
	\caption{
		Model luminosity in lines of \heii\,4686\,\AA\ ({\em thick line}), \ha\     
	    ({\em thin line}) and \heii\,304\,\AA\ ({\em dashed line}) as a function of fragments temperature. {\em Horizontal line} shows the observational luminosity 
	    of \heii\,4686\,\AA, {\em vertical line} corresponds to the temperature 
	    fixed by the description of the \heii\,4686\,\AA\ luminosity, the ratio 
	    of \heii\,4686\,\AA/\ha\ $\sim 5$ and assuming minimal luminosity of \heii\,304\,\AA.
	    	}
	\label{fig:excit}
\end{figure}

%==================================================================
\subsection{Results}  

The calculated luminosity vs. temperature for \ha, \heii\,4686\,\AA, and 
  \heii\,304\,\AA\ in SN~2020jfo at 2.1\,d after the explosion is shown in Fig.~2.
Input model parameters are listed in the first three lines of Table.
Lower three lines of Table contain the mass of the emitting gas, the parameter $\zeta$ and 
 the surface barion density determined by the total mass of fragments and their total surface area.
The mass and $\zeta$  are inferred from the agreement between the calculated and observational 
  luminosity of the 4686\,\AA\ line.
The parameter $\zeta$ is responsible for the choice of the optimal fragment surface 
  density that secures an acceptable value of the photon escape probability.
 
The collisional excitation of \heii\ results in the significant excess of the \heii\,4686\,\AA\   
  luminosity over that of \ha\ in the temperature range of $(4.5-8)\times10^4$\,K thus confirming 
  the above suggestion.
Figure~2 displays also the \heii\,304\,\AA\ luminosity; this line dominates in radiative energy losses in the considered temperature range and therefore is crucial for the energy balance.
   
A key modelling result is the fixing of the optimal temperature $T \approx 52000$\,K 
  at which three major properties of the \heii\,4686\,\AA\ emission are reproduced, namely, 
  the luminosity, the large flux ratio 4686\,\AA/\ha\ $\approx 5$, and the significant line optical depth  $\tau_{34} \approx 5$.
At this temperature the radiative energy losses are minimal  with the \heii\,304\,\AA\ luminosity of $4\times10^{40}$\ergs.

\section{Fragments heating}

The X-ray luminosity of the forward shock amounts to $10^{41}$\ergs at the considered stage (Utrobin \& Chugai 2024).
However, given the high temperature of the bremsstrahlung radiation ($\sim$300\,keV) and the low 
  surface density of fragments ($\sim$10$^{-4}$\gcmsq), the power absorbed by fragments turns 
  out mediocre $\lesssim$10$^{37}$\ergs, significantly smaller than the \heii\,304\,\AA\ luminosity.

An alternative mechanism is the electron thermal conductivity.
Let us evaluate this possibility neglecting at, the moment, the magnetic field.
Given the low cross section of the fast electron scattering and the 
 low surface density of fragments, the  scattering probability with a significant 
    energy transfer turns out small, $p \sim 10^{-3}$.
Maximal velocities of hot electrons are subrelativistic ($\beta = v/c \sim 1$), so 
  one needs to use the relativistic expression for the electron kinetic energy 
   $E = mc^2(\gamma -1)$, where $\gamma = 1/\sqrt{1-\beta^2}$. This expression enters the Maxwell distribution $f(\beta) \propto \beta^2 \exp{(-E(\beta)/kT)}$ with $\beta < 1$.
   
Line-emitting fragments reside in the thin layer close to the contact surface 
 SN ejecta/stellar wind ($R = 1$).
According to self-similar solution (Chevalier 1982a; Nadyozhin 1985), for 
the wind density  $\rho \propto 1/r^2$ and the SN density $\rho \propto 1/v^7$, at the fiducial radius $R = 1.05$ density  is  
  twice as high and temperature is twice as low as values at the shock front.
We adopt conditions in the hot plasma at the radius $R = 1.05$ 
  for the fragments environment.
In this case the average energy of hot electrons in the flux toward certain direction 
 is 164\,keV, while the average speed of electrons is $\beta = 0.54$.

The energy loss by a fast electron in plasma per unit length (Breizman et al. 2019) is 
\begin{equation}
\frac{dE}{ds} = -\frac{2\pi e^4n_e}{mc^2\beta^2}\ln{\left[\frac{m^2c^4(\gamma^2 - 1)(\gamma - 1)}{2(\hbar \omega_p)^2\gamma^2}\right]}\,,
\end{equation}
where $n_e$ is the electron number density in fragments, $m$ is the electron mass,
 $\omega_p$ is the plasma frequency, while other notations are standard.
The mean pass length of electron crossing flat fragment is 
  $l = 2b \approx 2.5\times10^4$\,cm, where $b = N_b/n_b \approx 1.25\times10^4$\,cm is 
  the average thickness of fragment.
The average energy lost by a fast electron during crossing fragment is  
 $\Delta E = (dE/ds)l$, while the energy flux deposited in a fragment is 
  $q = (1/4)y_en_sc\beta \Delta E \approx 9.7\times10^9$\ergs\,cm$^{-2}$, where 
  $y_e = 0.85$ is the number of electrons per baryon, $n_s = 4\times10^9$\cmq\ is 
  the number density of baryons at the shock front.
The total heating power for all fragments is then $L_{inj} = 8\pi r^2\zeta q \approx 6.2\times10^{40}$\ergs\ (heat flux through both sides is included).
The estimated power, therefore, fully compensates the energy loss via 
 the \heii\,304\,\AA\ emission ($4\times10^{40}$\ergs).
 
The magnetic field that can be generated by turbulence related to the RT instability 
 (Chevalier 1982b) unlikely inhibits the hot electrons penetration into fragments.
 The giroradius of a fast electron $r_e = \beta \gamma mc^2/eB$ is equal to the fragment  
    thickness $b \approx 1.2\times10^4$\,cm for the magnetic field of   
   $B \approx 0.3$\,G.
This value is comparable to the field invoked for the interpretation of early radio emission of SNe~IIP 
 (Chevalier et al. 2006; Yadav et al. 2014). 
We therefore do not expect the magnetic field in SN~2020jfo substantially exceeding  0.3\,G.   
Unexpectedly, the effect of such a field is boosting, and not inhibiting, the heating by electrons, since the electron pass curvature caused by the magnetic field  
  increases the residence time of the fast electron in fragments.

\section{Discussion}

The proposed model for the bfoad \heii\,4686\,\AA\ emission in 
the early spectrum of SN~2020jfo explains satisfactorily its major properties   --- large blueshift, 
  the line luminosity, the large flux ratio \heii\,4686\,\AA/\ha\, and the significant line optical depth.
The model thus catches the basic physics involving the line-emitting 
  dense fragments with the temperature of $\approx$5$\times10^4$\,K embedded in the hot gas of the  forward shock.
The proposed model is applicable to other SNe~IIP with the broad \heii\,4686\,\AA\ emission in the early spectra.
This  model is an alternative with respect to the model of the early spectra  of SNe~IIP (Dessart et al. 2017) without the adiabatic forward shock, 
 but showing the broad \heii\,4686\,\AA\ emission.

The line-emitting gas responsible for the broad 4686\,\AA\ emission in our model is    
 related to the boundary thin dense shell that is liable to the RT instability.  
This boundary thin shell forms at the SBO stage of the exploding SN~IIP   
  (Grasberg et al. 1971; Chevalier 1981); a similar thin shell with small mass  
  ($\sim$10$^{-6}$\msun) is present in the hydrodynamic model of SN~2020jfo (Utrobin \& Chugai 2024).
Three hours later after the SBO, the interaction with the  wind increases the 
 thin shell mass up to $\sim$10$^{-5}$\msun, while  the shell density 
 contrast with respect to the preshock wind amounts to $\sim$10$^4$.   
The heavy deceleration of the thin shell  suggests the significant  effective acceleration $g = -d^2R/dt^2 \sim 2\times10^3$\,cm\,s$^{-2}$, while the high density contrast ($\rho_2/\rho_1 \sim 10^4$)  implies the large Atwood number $A = (\rho_2-\rho_1)/(\rho_1+\rho_2) \sim 10^4$. 
The large $g$ and $A$ result in the exponential growth of the RT instability with the increment $\gamma \approx \sqrt{Agk} \sim  
3\times10^{-3}/\lambda_{13}$\,s$^{-1}$, where $\lambda = 2\pi/k$ and  $\lambda_{13}$ 
 is in units of $10^{13}$\,cm.  

Perturbations in the form of spikes of dense gas penetrate the hot rarefied gas; the latter, 
 in turn, forms bubbles that decelerate the dense shell between spikes.  
In the non-linear stage the exponential growth is replaced by the spike 
 growth as
  $h \approx Agt^2$ (Fermi \& von Neumann 1953).   
In fact, the spike growth is limited by the KH instability that
results in the spike stripping and the formation of mushroom structures which are
 apparent in 3D numerical simulations (e.g., Blondin \& Ellison 2001).
As a result, the KN instability brings about an ensemble of corrugated sheets of dense line-emitting fragments 
  embedded in the rarefied hot gas of the forward shock.   
The collateral turbulence results in the progressive fragmentation and 
   mixing of fragments. 
  Eventually, dense fragments dissolve and the broad \heii\,4686\,\AA\ emission disappears.
This scenario permits us to understand the early emergence  and a brief lifetime of the broad \heii\,4686\,\AA\ emission. 
 
An interesting outcome of the proposed model for the   
  early broad  emission of the \heii\,4686\,\AA\ line is an expected flash of 
   a broad emission of the \ovi\,1032,\,1038\,\AA\ resonance doublet 
  coeval with the fading 4686\,\AA\ line.
  The point is that the advanced fragmentation results in the increase of fragments total surface area, which brings about a more efficient heating that 
  would favor the high emissivity of the \ovi\ resonant doublet at the temperature of $\sim$10$^5$\,K.
The predicted 
\ovi\,1032,\,1038\,\AA\ emission would have the same width as  
 the broad 4686\,\AA\ line, in contrast to lines with narrow core and broad  wings originating from the photoionized preshock wind (Groh 2014).

\section{Conclusions}

Below is a brief r\'{e}sum\'{e} of major results.

\begin{itemize}
	\item 
	We propose the model of the broad \heii\,4686\,\AA\ emission in the early spectrum
	of SN~2020jfo and the scenario that accounts for the early emergence and 
	 a brief duration of this emission.
	\item
   The broad 4686\,\AA\ line is emitted by fragments of the boundary thin dense shell 
    embedded in a hot gas of the adiabatic forward shock. 
    The shell fragmentation and the subsequent fragments mixing with hot gas is the outcome of the RT instability.	
	\item 
	Calculations of the ionization and excitation of helium and hydrogen reproduce the 
	\heii\,4686\,\AA\ luminosity, the large flux ratio 4686\,\AA/\ha, and a
	significant optical depth of the 4686\,\AA\ line.
		\item	
			It is shown that the fragment heating can be produced by hot electrons of the 
	forward shock.

\end{itemize}

\clearpage

\section{References}

\noindent
\medskip 
Andrews J. E., Sand D. J., Valenti S. et al., Astrophys. J. {\bf 885}, 43 (2019)\\
\medskip
Blinnikov S. I,, AIP Conference Proceedings {\bf 1016}, 241 (2008)\\
\medskip
Blinnikov S. I., Bartunov O. S., Astron. Astrophys. {\bf 273}, 106 (1993)\\
\medskip 
Blondin J. M., Ellison D. C., Astrophys. J. {\bf 560}, 244 (2001)\\
\medskip
Bohigas J., Astrophys. J. {\bf 674}, 954 (2022)\\
\medskip
Breizman B. N., Aleynikov P., Hollmann E. M., Lehnen M.,
Nuclear Fusion {\bf 59}, Issue 8, article id. 083001 (2019)\\
\medskip 
Bullivant C. et al., Monthly Not. R. Astron. Soc. {\bf 476}, 1497 (2018)\\
\medskip 
Chevalier R. A., Fransson C., Nymark T. K.,  Astrophys. J. {\bf 641}, 1029 (2006)\\ 
\medskip 
Chevalier R. A., Astrophys. J. {\bf 258}, 790 (1982a)\\
\medskip 
Chevalier R. A., Astrophys. J. {\bf 259}, 302 (1982b)\\
\medskip
Chevalier R. A., Fundamentals Cosmic Phys., {\bf 7}, 1 (1981)\\
\medskip
Chugai N. N., Monthly Not. R. Astron. Soc. {\bf 494}, L86 (2020)\\
\medskip
Chugai N. N., Blinnikov S. I., Fassia A. et al., Monthly Not. R. Astron. Soc.  
{\bf 330}, 473 (2002)\\
\medskip
Chugai N. N., Monthly Not. R. Astron. Soc. {\bf 326}, 1448 (2001)\\
\medskip 
Dessart L., Hillier D. J., Audit E., Astron. Astrophys. {\bf 603A}, 51 (2017)\\
\medskip
Fermi E., von Neumann J., Technical Report no. AECU-2979,
Los Alamos Scientific Laboratory, (OSTI ID: 4373391) (1953) \\
\medskip 
Grasberg E. K., Imshennik V. S.,  Nadyozhin D. K., Astrophys. Space Sci. {\bf 10}, 3 (1971)\\
\medskip
Grefenstette B. W., Brightman M., Earnshaw H. P., Harrison F. A., Margutti R.,
Astrophys. J. {\bf 952}, id. L3, 6 pp. (2023) \\
\medskip
Groh J. H., Astron. Astrophys. {\bf 572}, L11 (2014)\\
\medskip 
Jacobson-Gal\'{a}n W. V.,  Dessart L., Margutti R. et al., Astrophys. J. {\bf 954}, L42 (2023)\\
\medskip
Morozova V.,  Piro A. L., Valenti S, Astrophys. J. {\bf 838}, 28 (2017)\\
\medskip
Nadyozhin D. K., Astrophys. Spce Sci. {\bf 112}, 225 (1985)\\
\medskip 
Osterbrock D. E., Ferland G. J., {\em Astrophysics of gaseous nebulae and active galactic nuclei} USA: University Science Books, 2006\\ 
\medskip
Quimby R. M.,  Wheeler J. C., H\"{o}flich P. et al., Astrophys. J. {\bf 666}, 1093 (2007)\\
\medskip 
Shrestha M., Pearson J.,  Wyatt S. et al., eprint arXiv:2310.00162 (2023)\\
\medskip
Teja R. S., Singh A., Sahu D. K. et al.,  Astrophys. J. {\bf 930}, 34 (2022)\\
\medskip 
Utrobin V. P., Chugai N. N.,  Monthly Not. R. Astron. Soc. {\bf 527}, 6227  (2024)\\
\medskip
Vernazza J. E., Avrett E. H., Loeser R., 
Astrophys. J. Suppl. Ser. {\bf 45}, 635 (1981) \\
\medskip	
van Regemorter H., Astrophys. J. {\bf 136}, 906 (1962)\\ 
\medskip 
Yadav N., Ray A., Chakraborti S. et al.,  Astrophys. J. {\bf 782}, 30 (2014)\\
\medskip
Yaron O.,  Perley D. A., Gal-Yam A. et al., Nature Physics {\bf 13}, 510 (2017)\\
\clearpage

\end{document}